\documentclass[letterpaper, 10 pt, conference]{ieeeconf}  % Comment this line out if you need a4paper

\IEEEoverridecommandlockouts                              % This command is only needed if 
\usepackage{graphics} % for pdf, bitmapped graphics files
\usepackage{epsfig} % for postscript graphics files
\usepackage{mathptmx} % assumes new font selection scheme installed
\usepackage{times} % assumes new font selection scheme installed
\usepackage{amsmath} % assumes amsmath package installed
\usepackage{amssymb}  % assumes amsmath package installed
\usepackage{multirow}
\usepackage[margin=19.1mm]{geometry}
\usepackage{stfloats} % to correctly put table at bottom

\usepackage{ctable} % for \specialrule command

\title{\LARGE \bf
Deep Feature Fusion via Graph Convolutional Network for Intracranial Artery Labeling
}

\author{Yaxin Zhu$^{*1}$, Peisheng Qian$^{*2}$, Ziyuan Zhao$^{2}$, and Zeng Zeng$^{\dagger2}$% <-this % stops a space
\thanks{$^*$Contributed equally. $^{\dagger}$Corresponding author, email: zengz@i2r.a-star.edu.sg.
$^{1}$School of Microelectronics, Shanghai University, Shanghai, China, 200444. $^{2}$Institute for Infocomm Research, A*STAR, 1 Fusionopolis Way, 21-01 Connexis, Singapore 138632.}% <-this % stops a space
}

\begin{document}

\maketitle
\thispagestyle{empty}
\pagestyle{empty}

%%%%%%%%%%%%%%%%%%%%%%%%%%%%%%%%%%%%%%%%%%%%%%%%%%%%%%%%%%%%%%%%%%%%
\begin{abstract}
Intracranial arteries are critical blood vessels that supply the brain with oxygenated blood. Intracranial artery labels provide valuable guidance and navigation to numerous clinical applications and disease diagnoses. Various machine learning algorithms have been carried out for automation in the anatomical labeling of cerebral arteries. However, the task remains challenging because of the high complexity and variations of intracranial arteries. This study investigates a novel graph convolutional neural network with deep feature fusion for cerebral artery labeling. We introduce stacked graph convolutions in an encoder-core-decoder architecture, extracting high-level representations from graph nodes and their neighbors. Furthermore, we efficiently aggregate intermediate features from different hierarchies to enhance the proposed model's representation capability and labeling performance. We perform extensive experiments on public datasets, in which the results prove the superiority of our approach over baselines by a clear margin. 
\newline
\indent \textit{Clinical relevance}— The graph convolutions and feature fusion in our approach effectively extract graph information, which provides more accurate intracranial artery label predictions than existing methods and better facilitates medical research and disease diagnosis.
\end{abstract}

%%%%%%%%%%%%%%%%%%%%%%%%%%%%%%%%%%%%%%%%%%%%%%%%%%%%%%%%%%%%%%%%%%%%%%%%%%%%%%%%
\section{INTRODUCTION}
% magnetic resonance angiography 磁共振血管造影
Intracranial arteries (ICA) play a crucial role in supplying human brains with sufficient blood. Disease in intracranial arteries,~\emph{e.g.,} cerebral aneurysm and cerebral infarction, severely harms the patients' health and significantly lowers the survival rate. Accurate labeling of ICA provides deep insights into brain vascular structures, guiding subsequent clinical diagnosis and research on brain magnetic resonance imaging (MRI) and magnetic resonance angiography (MRA) images. Manual labeling of ICA demands a considerable amount of time and expertise. Therefore automation of ICA labeling becomes imminent.

 Continuous efforts have been made in the automation of ICA labeling. Handcrafted and statistical features are widely used in early research, including the vessel count (VC), average radius (AVRAD), sum of angles metric (SOAM), volumetric flow rate (VFR), etc.~\cite{bullitt2005vessel, Geri2017, Wolterink2019}. Morphological constraints of predefined lengths and volumes of the arteries are also taken as classification criteria in~\cite{dunas2015}. However, the problem of ICA labeling remains challenging because of the high complexity in cerebral arteries and high variability in the vascular structure. For instance, it is claimed that the presence of partial integral circle of Willis (CoW) is higher than $70\%$ in healthy subjects, while non-variation CoW integrity is only found in $7.57\%$ of the investigated subjects~\cite{qiu2015mra}. Moreover, imaging quality also limits the labeling accuracy, in which crossing and overlapping arteries are indistinct. 

Recently, graph representation and graph feature learning approaches have been studied on large datasets with heterogeneous sources~\cite{chen2019gated,qu2020using,zeng2021robust,qu2020multi}. A directed graph representation of intracranial arteries is proven effective in~\cite{robben2016simultaneous}, in which the ICA network is abstracted as a graph of connected 3D points and centerlines of arteries. The emerging graph neural network (GNN) is employed to learn informative properties of arteries for ICA labeling~\cite{chen2020automated}. Despite the progress from handcrafted features, these methods either suffer from high model complexity~\cite{robben2016simultaneous} or limited feature extraction capability~\cite{chen2020automated}.

\begin{figure}[t]
    \centering
    \includegraphics[width = 5cm]{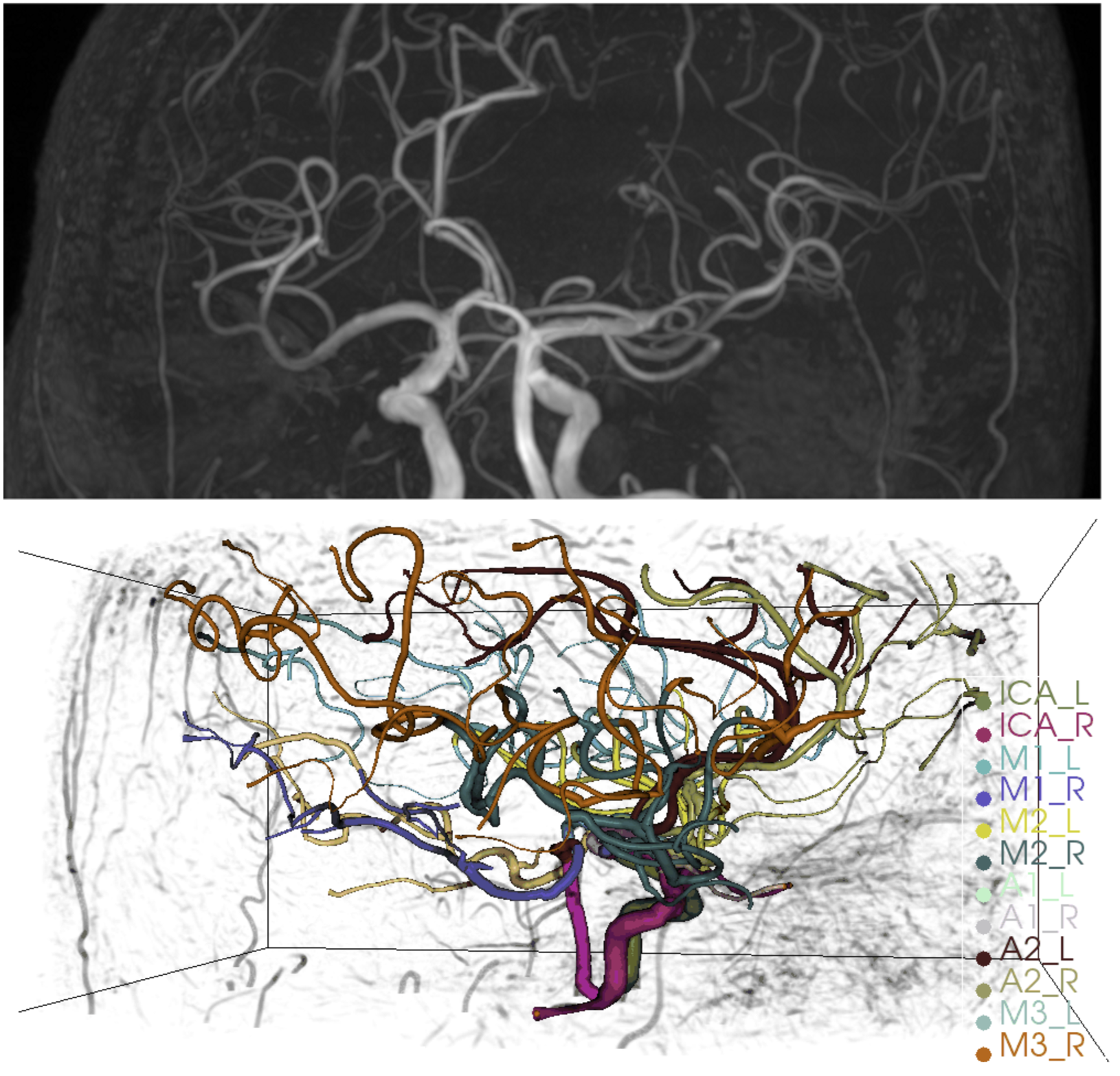}
    \caption{Top: A Magnetic Resonance Angiography (MRA) scan of cerebral arteries. Bottom: Intracranial arteries labeled in different colors. Intracranial arteries are labeled and visualized in the 3D vessel tracing software developed by~\cite{Chen2018}.}
    \label{fig:fig_1}
\end{figure}

In this study, we propose a graph convolutional network with deep feature fusion to enhance the automated feature learning from graph embeddings of vascular information for ICA labeling. The graph convolution operations effectively exploit structural information of the current node and its direct and indirect neighbors in a graph. The stacked graph convolutional layers leverage properties from local to global scales in the graph to determine the label of the current node. We further enhance the feature representation by aggregating hidden features from intermediate levels of the network with concatenation and pooling operations. Contributions in this paper are summarized as follows: 
\begin{figure*}[t]
\centering
\includegraphics[width=0.8\textwidth]{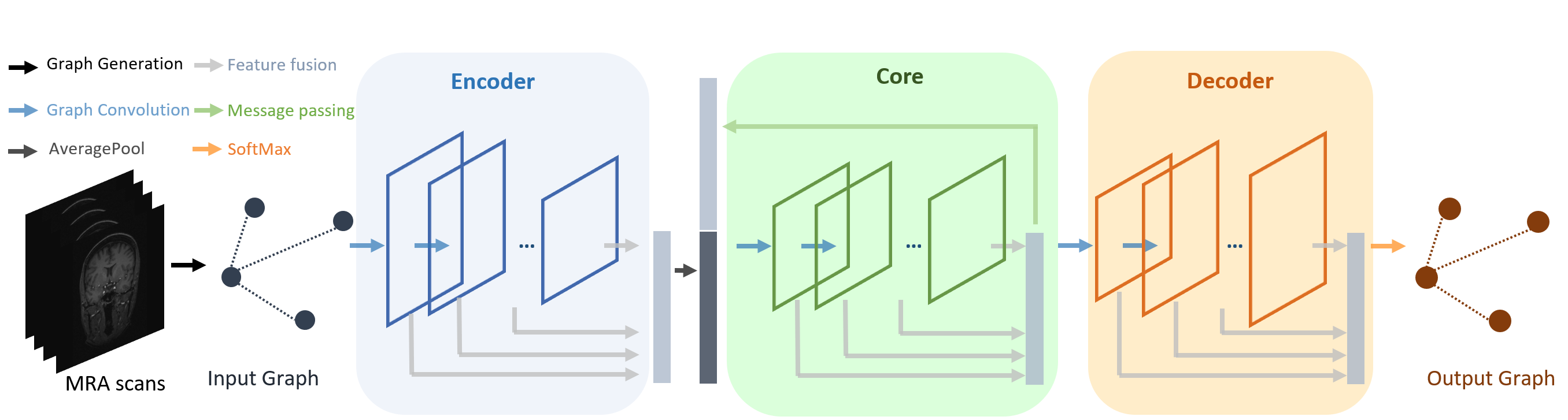}
\caption{The proposed network architecture for ICA labeling. Node features (bifurcation/ending points) and edge features (arteries) are taken as input in a graph structure generated from MRA scans. The output graph contains updated node and edge features and their respective ICA labels.}
\label{fig:arch}
% \vspace{-15pt}
\end{figure*}

\begin{itemize}
    \item We develop a novel graph convolutional network with deep feature fusion for ICA labeling. The model captures artery structures and relations at different scales, based on which the artery labels are determined.
    \item We perform extensive experiments of our approach on the newly released ICA dataset~\cite{chen2020automated}. Extensive experimental results show noticeable improvement in our method compared to baselines.
\end{itemize}

% Morphological variations
\section{RELATED WORK}
Early studies on ICA labeling rely on handcrafted features and templates generalized from small datasets. Abnormalities in vessel tortuosity associated with brain tumors are characterized in~\cite{bullitt2005vessel}. A CoW template is generalized from $5$ subjects, and $15$ scans are aligned for ICA labeling~\cite{takemura2006automatic}. A probabilistic atlas over cerebral arteries is built from $46$ individuals, from which an automatic atlas-based artery identification algorithm (AAIM) is derived~\cite{dunas2015}. These studies are bounded by the limited sizes of datasets, while there are much more variations in vascular structures unrepresented. 

Graph representations of arteries are adopted in recent research, based on which Bilgel~\emph{et al.} develop a random forest classifier for artery labeling and a Bayesian network incorporating messages from connected vessels~\cite{bilgel2013automated}. Graph templates are used together with maximizing a posteriori probability (MAP) estimate for CoW labeling~\cite{bogunovic2013anatomical}. Simultaneous segmentation and labeling of cerebral vasculature are carried out, in which anatomical labels are determined by a branch-and-cut algorithm on overfull graphs~\cite{robben2016simultaneous}. 

A graph neural network (GNN) is designed to exploit global artery structures and relations in~\cite{chen2020automated}. Chen~\emph{et al.} also release an ICA dataset with heterogeneous sources and the construction from MRA to graph embeddings. While the research in~\cite{chen2020automated} constitutes the baselines of our work, we argue that substantial improvement can be made to the GNN architecture in~\cite{chen2020automated}, in which the multi-layer-perception (MLP) based model is insufficient to utilize information from the graph fully. Convolutional neural network (CNN) have obtained great success on various biomedical applications~\cite{litjens2017survey,gu2020multi,zhao2021mt,qian2021two}. And Graph convolutional neural network (GCN) generalizes classic CNN on graph structure~\cite{wu2020comprehensive}. An efficient GCN has been implemented to obtain graph node features~\cite{kipf2017semi}, which inspires our approach. 

\section{METHODOLOGY}
\label{sec:method}
The proposed network is illustrated in Fig.~\ref{fig:arch}, in which ICA graphs containing node and edge features are taken as input, respectively. The network consists of $3$ components sequentially,~\emph{i.e.}, encoder, core, and decoder. Graph convolutional layers are stacked in each component to enable feature extraction from graph nodes and their neighbors, which is described in Sec.~\ref{sec:graph_convolution}. Deep feature fusion is carried out on output features from each component to enhance the latent representation further, which is elaborated in Sec.~\ref{sec:feature_fusion}. Predictions of node and edge types are computed in the last fully connected layer based on restored node and edge features in the decoder and embedded in the output graphs for nodes and edges. 

\subsection{Graph Convolutional Network}
\label{sec:graph_convolution}
The graph convolution is the essential operation that extracts node features based on their structural information. The $3$ components of our network consist of stacked graph convolutional layers with Sigmoid activation ($\sigma$), as illustrated in Fig.~\ref{fig:arch}. Let $A\in R^{N\times N}$ denote the normalized adjacency matrix, $X_i\in R^{N\times D}$ denote the input signals of the $i$th graph convolutional layer, $X_{i+1}\in R^{N\times M}$ denote the output, and $W_i\in R^{D\times M}$ stand for the parameter matrix of the $i$th graph convolution operation. A graph convolutional layer in our work is defined as follows,
\begin{equation}
    X_{i+1}=\sigma(A\times X_i\times W_i)
    \label{eqn:graph_convolution},
\end{equation}
where $i=1,2,3,...$ depending on the depth of the network. In each graph convolutional layer, attributes in the edge graph are updated first, followed by attributes in the node graph.

\subsection{Deep Feature Fusion}
\label{sec:feature_fusion}
We adopt $2$ strategies to aggregate latent features for model optimization. First, we concatenate intermediate features from graph convolutional layers in the encoder as the input for the following component before the average pooling operation. The updated feature covers high-resolution, low-context information from the early feature extraction stages and low-resolution, high-context information from the later stages. The feature fusion operation is represented in Eqn.~\ref{eqn:fusion},
\begin{equation}
    X_{fusion}=\bigoplus AvgPool(X_i),
    \label{eqn:fusion}
\end{equation}
where the maximum $i$ depends on the number of graph convolutional layers in the encoder. Features are directly concatenated in the core and decoder, in which $X_{fusion}$ is calculated as follows,
\begin{equation}
    X_{fusion}=\bigoplus X_i
    \label{eqn:fusion_no_pool}.
\end{equation}

Furthermore, we iteratively update feature representations in the core component by concatenating its output to output from the encoder as its input for the next training iteration of the core. We follow the notation in~\cite{chen2020automated} and refer to this process as message passing. In the first round of message passing, the core takes duplicated output from the encoder. In the subsequent iterations, it takes its output in the previous iteration concatenated to output from the encoder as input. In our experiments, message passing is repeated for $10$ times.

\section{EXPERIMENTS}
\label{sec:experiments}

\subsection{Dataset and Implementation}
\label{sec:dataset_implementation}
We carry out experiments on the ICA dataset released in~\cite{chen2020automated}. Both healthy subjects and subjects with vascular diseases are included in the dataset. The dataset is randomly split into training, validation, and test sets. Following~\cite{chen2020automated}, MRA images with ICA labels are pre-processed to generate relational graphs with attributes. In a Graph $G=(V, E)$, $V=\{v_i\}$ stands for all starting and ending points of artery centerlines with node features. $E=\{e_k, r_k, s_k\}$ denotes all edges where the edge $k$ is connected to nodes $r_k$ and $s_k$. Features for node $v_i$ include the 3D coordinates, radius, and approximated directional embedding. Features for edge $e_k$ include the edge direction, distance, and mean radius. There are $21$ bifurcation or ending types of nodes, and the edges are labeled by their starting and ending nodes.

In our implementation, node positions are normalized in images with different resolutions. Random translation of node position up to $10\%$ is applied as data augmentation. The network is randomly initialized. The Adam optimizer is used to train the network with initial learning rate of $0.001$ and a batch size of $32$ for $12000$ epochs. The model is evaluated during training on the validation set, and the epoch with the best validation result is used for testing.

\begin{table*}[t]
\caption{Experimental results on the test dataset. $\uparrow$ indicates that higher values are better. $\downarrow$ indicates that lower values are better. $-$ stands for results that are not available.}
\label{tab:result}
\centering
\begin{tabular}{c | c c c c c c c}
\specialrule{.1em}{.05em}{.05em} 
Method & Node\_Acc$\uparrow$  & Node\_Wrong$\downarrow$ & CoW\_Node\_Solve$\uparrow$ & Edge\_Acc$\uparrow$ & Precision$\uparrow$ & Recall$\uparrow$\\
\hline
Template~\cite{Geri2017} & $0.7316$ &$31.6$ & $0.0476$ & $0.7934$ & - & -\\
\hline
Atlas~\cite{dunas2015} & $0.8856$ &  $13.5$ & $0.0095$ & $0.7010$ & - & - \\
\hline
MAP~\cite{Chen2019} & $0.9153$ & $10.0$ & $0.0476$ & $0.3304$ & - & -  \\
\hline
GNN~\cite{chen2020automated} &$0.9746$ & $3.0$ & $0.6381$ & $0.9246$ & $0.9130$ & $0.8381$  \\
\specialrule{.1em}{.05em}{.05em} 
\textbf{GCN} & 0.9739 & 2.3 & \textbf{0.6667} & 0.9258 & 0.9323 & 0.8667 \\
\hline
\textbf{GCN\_Fusion} & \textbf{0.9777} & \textbf{1.9} & 0.6190 & \textbf{0.9336} & \textbf{0.9464} & \textbf{0.8836} \\
\specialrule{.1em}{.05em}{.05em} 
\end{tabular}
\end{table*}
\subsection{Evaluation Metrics}
Following previous works~\cite{dunas2015, chen2020automated}, we evaluate our method on the following metrics:
\begin{itemize}
    \item Node\_Acc: The accuracy of node label prediction.
    \item Node\_Wrong: The average number of wrong node predictions in $1$ MRA scan.
    \item CoW\_Node\_Solve: The percentage of scans in which all CoW nodes are correctly predicted.
    \item Edge\_Acc: The accuracy of edge label prediction.
    \item Precision: The overall precision for both node and edge prediction.
    \item Recall: The overall recall for both node and edge prediction.
\end{itemize}

\subsection{Baseline Methods}
The following baselines are studied. In~\cite{Chen2018}, a probability model is developed to label ICA based on hand-crafted features. More statistical patterns are exploited to label vessels in~\cite{Geri2017}. An atlas-based artery identification method is studied in~\cite{dunas2015}. GNN with hierarchical refinement is probed in~\cite{chen2020automated}, which constitutes the backbones of our approach. Our methods with different configurations are described in the following, 
\begin{itemize}
\item GCN: Graph convolutional network.
\item GCN\_Fusion: Graph convolutional network with deep feature fusion.
\end{itemize}
\subsection{Results and Discussion}
\label{sec:results}
We compare experimental results of baselines and our approach, as presented in Table.~\ref{tab:result}. It is evident that our method outperforms baselines in all evaluation metrics. We can observe that GNN~\cite{chen2020automated} has made remarkable improvements from other baselines on ICA data with a graph structure, which is further enhanced in our paper by substituting the multi-layer-perceptron architecture in~\cite{chen2020automated} with graph convolutions and deep feature fusion. It is worth mentioning that there is a heavily engineered hierarchical refinement postprocessing step after GNN in~\cite{chen2020automated}, which is also incorporated in our method.

Model performance comparison on posterior cerebral arteries is shown in Table.~\ref{tab:per_cls}, in which our model consistently outperforms the baseline except in the Pcomm/ICA nodes. There is a higher mutation rate in posterior cerebral arteries than the rest of the arteries, which increases its structural variability and makes it more challenging to label. Therefore the results confirm the robustness of our method on challenging cases. We also perform ablation studies on the model architecture. As revealed in Table.~\ref{tab:ablation}, deeper GCN networks generally perform better than shallower ones. Besides, the average pooling operation proves to be an essential step in deep feature fusion, which provides a noticeable increment in all evaluation metrics.
\begin{table}[ht]
\caption{Precision and recall of prediction in posterior cerebral arteries and occipital arteries.}
\label{tab:per_cls}
\centering
\begin{tabular}{c|cccc}
\specialrule{.1em}{.05em}{.05em} 
\multicolumn{1}{c}{} & \multicolumn{2}{c}{GNN~\cite{chen2020automated}}& \multicolumn{2}{c}{GNN\_Fusion} \\
\hline
Node & Precision & Recall & Precision & Recall\\
\hline
 OA\_L & 0.9828 & 0.9344 & \textbf{0.9831} & \textbf{0.9508} \\
\hline
 OA\_R & 0.9643 & 0.9000 & \textbf{0.9818} & 0.9000\\
\hline
PCA/BA & 0.8447 & 0.8878 & \textbf{0.8654} & \textbf{0.9184} \\
\hline
P1/P2/Pcomm\_L & 0.7455 & 0.7593 & \textbf{0.8276} & \textbf{0.8889} \\
\hline
P1/P2/Pcomm\_R & 0.7407 & 0.7843 & \textbf{0.8800} & \textbf{0.8627} \\
\hline
Pcomm/ICA\_L & \textbf{0.9070} & 0.7800 & 0.8958 & \textbf{0.8600} \\
\hline
Pcomm/ICA\_R & \textbf{0.9091} & \textbf{0.8163} & 0.9048 & 0.7755 \\
\hline
\end{tabular}
\end{table}
\begin{table}[ht]
\caption{Experimental results on the exclusion of model components in GCN with deep feature fusion.}
\label{tab:ablation}
\centering
\begin{tabular}{c|ccc}
\specialrule{.1em}{.05em}{.05em} 
Experiments & Node\_Acc & Precision &Recall \\
\hline
1 GCN layer & 0.9700 & 0.9349 & 0.8499  \\
\hline
2 GCN layers & 0.9731 & 0.9374 & 0.8659 \\
\hline
3 GCN layers & 0.9615 & 0.9101 & 0.8242  \\
\hline
4 GCN layers & 0.9715 & 0.9373 & 0.8601 \\
\hline
GCN\_Fusion, no pooling & 0.9739 & 0.9323 & 0.8667  \\ 
\hline
GCN\_Fusion & \textbf{0.9777} & \textbf{0.9464} & \textbf{0.8836}\\
\hline
\end{tabular}
\end{table}

\begin{figure}[ht]
    \centering
    \includegraphics[width = 8cm]{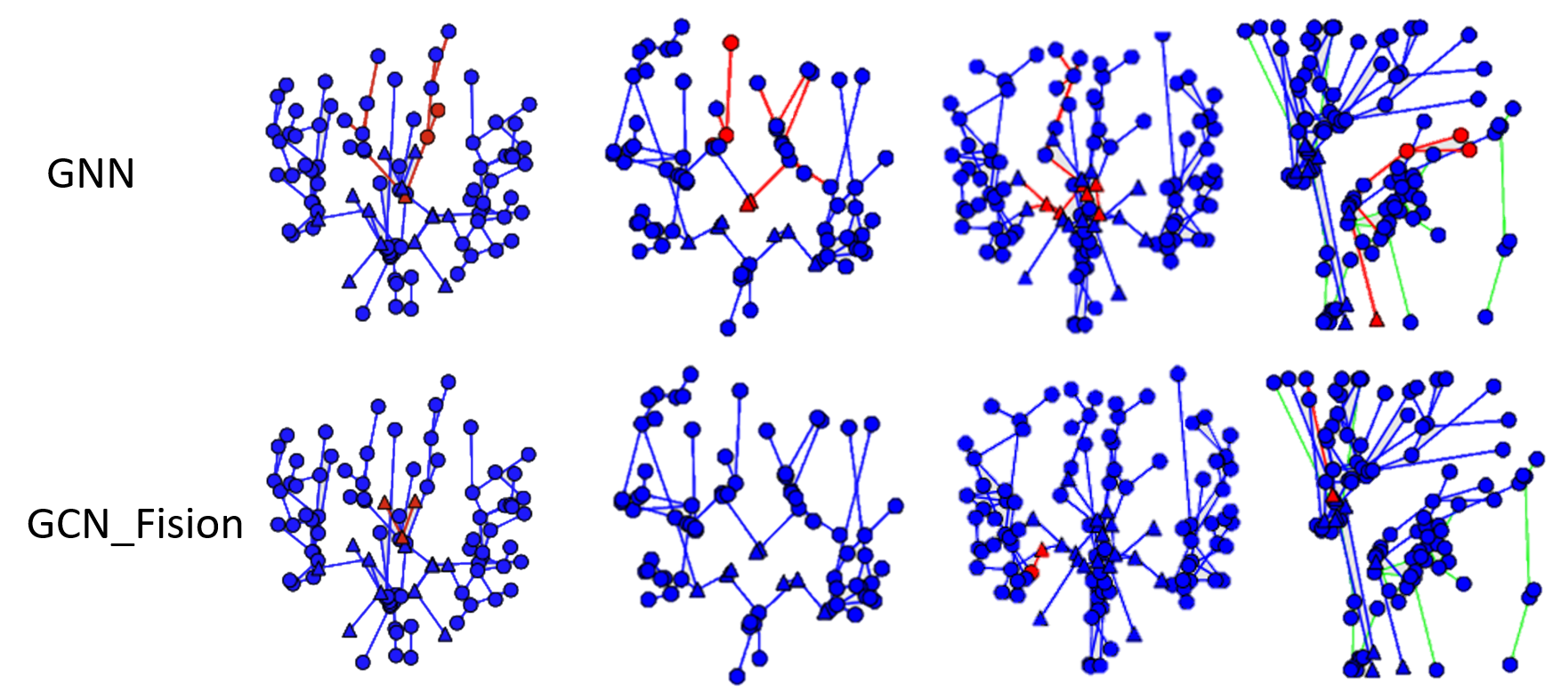}
\caption{Qualitative comparison between visualized predictions from GNN~\cite{chen2020automated} and our GCN\_Fusion. The blue color indicates correct prediction. The red color indicates existent edges with wrong labels. The green color indicates non-existent edges. Circular dots represent nodes outside CoW. Triangular dots represent nodes within CoW.}
    \label{fig:visualization}
% \vspace{-15pt}
\end{figure}

Inference results on the test set are illustrated in Fig.~\ref{fig:visualization}. In the first $3$ column, we can observe fewer prediction errors from GCN\_Fusion than the baseline. However, similar mistakes are made in the last column, where both models predict non-existent edges at the same positions. Future studies could investigate these cases.
\section{CONCLUSIONS}
In a nutshell, we design a network with deep feature fusion for intracranial artery labeling, which consists of stacked graph convolutions in the encoder-core-decoder architecture. Latent features are pooled in the encoder and concatenated in each network component to enhance node and edge representation for better labeling performance. Experimental results have validated our approach. In the future, we can conduct additional experiments on multiple independent datasets, and focus on incorrectly labeled arteries.

\addtolength{\textheight}{-12cm}   % This command serves to balance the column lengths

%%%%%%%%%%%%%%%%%%%%%%%%%%%%%%%%%%%%%%%%%%%%%%%%%%%%%%%%%%%%%%%%%%%%%%%%%%%%%%%

%%%%%%%%%%%%%%%%%%%%%%%%%%%%%%%%%%%%%%%%%%%%%%%%%%%%%%%%%%%%%%%%%%%%%%%%%%%%%%%%



% \section*{ACKNOWLEDGMENT}

%%%%%%%%%%%%%%%%%%%%%%%%%%%%%%%%%%%%%%%%%%%%%%%%%%%%%%%%%%%%%%%%%%%%%%%%%%%%%%%%

\bibliographystyle{IEEEbib}
\bibliography{refs.bib}

\end{document}